\documentclass[useAMS,usenatbib]{mn2e}
\usepackage{times}
\usepackage{url}
\usepackage{graphicx}
\usepackage[usenames]{color}
\DeclareGraphicsExtensions{.pdf,.png,.jpg,.mps,.eps,.ps}
\usepackage{amsmath}
\usepackage{natbib}
\usepackage{tablefootnote}
\usepackage{cancel}
\usepackage{soul}
\usepackage[normalem]{ulem}

\newcommand\funit{\rm \,erg\,cm^{-2}\,s^{-1}}

\newcommand\si{Si\,{\sc ii}\,}
\newcommand\Ni{Ni\,{\sc ii}\,}

\newcommand{\pasp}{PASP} \def\apj{ApJ} \def\mnras{MNRAS}
\def\aap{A\&A} \def\apjl{ApJ}  
 \def\apjs{ApJS}  
   \def\aaps{AAPS} \def\rmxaa{RMxAA}

\usepackage{graphicx}

\title{Ultraviolet emission lines of \si in cool star and solar spectra}

\author [Laha et al.] {Sibasish Laha$^1$\thanks{email: sib.laha@gmail.com}, Francis P. Keenan$^2$, Gary J. Ferland$^3$, Catherine A. Ramsbottom$^1$, \newauthor Kanti M. Aggarwal$^{2}$, Thomas R. Ayres$^4$, Marios Chatzikos$^3$, Peter A. M. van Hoof$\,^5$, \newauthor and Robin J. R. Williams$^6$\\ 
$^{1}$ Centre for Theoretical Atomic, Molecular and Optical Physics, School of Mathematics and Physics, 
Queen's University Belfast, Belfast BT7 1NN, Northern Ireland, U.K.\\
$^2$ Astrophysics Research Centre, School of Mathematics and Physics, Queen's University Belfast, Belfast BT7 1NN, Northern Ireland, U.K.\\
$^3$  Department of Physics and Astronomy, The University of Kentucky, Lexington, KY 40506, U.S.A. \\
$^4$ Center for Astrophysics and Space Astronomy, 389 UCB, University of Colorado, Boulder, CO 80309, U.S.A.\\
$^5$Royal Observatory of Belgium, Ringlaan 3, 1180 Brussels, Belgium \\
$^6$AWE plc, Aldermaston, Reading RG7 4PR, UK.}

\date{\today}
\begin{document}
\pagerange{\pageref{firstpage}--\pageref{lastpage}} \pubyear{2013}




\maketitle
\label{firstpage}
\begin{abstract}
Recent atomic physics calculations for \si are employed within the Cloudy modelling code
to analyse Hubble Space Telescope (HST) STIS ultraviolet spectra of three cool stars, $\beta$ Geminorum, $\alpha$ Centauri A and B, as well as previously published HST/GHRS observations of $\alpha$ Tau, plus solar quiet Sun data from the High Resolution Telescope and Spectrograph. Discrepancies found previously between theory and observation for line intensity ratios involving the 3s$^{2}$3p $^{2}$P$_{J}$--3s3p$^{2}$ $^{4}$P$_{J^{\prime}}$ 
intercombination multiplet of \si at $\sim$\,2335\,\AA\ are significantly reduced, as are those for ratios containing the 3s$^{2}$3p $^{2}$P$_{J}$--3s3p$^{2}$ $^{2}$D$_{J^{\prime}}$ transitions at $\sim$\,1816\,\AA. This is primarily due to the effect of the new \si transition probabilities. However, these atomic data are not only 
very different from previous calculations, but also show large disagreements with measurements, specifically 
those of Calamai et. al. (1993) for the intercombination lines. New measurements of transition probabilities for \si are hence urgently required to confirm (or otherwise) the accuracy of the recently calculated values. If the new calculations are confirmed, then a long-standing discrepancy between theory and observation will have finally been resolved. However, if the older measurements are found to be correct, then the agreement between theory and observation is simply a coincidence and the existing discrepancies remain. 

 \end{abstract}


\vspace{0.5cm}

\section{INTRODUCTION}

Emission lines involving transitions in Si\,{\sc ii} are important diagnostics for astrophysical plasmas,
with several intensity ratios being 
sensitive to temperature and density variations \citep{1985ApJ...289..844D,1991MNRAS.253..123J,1992ApJ...387..726K}. Over the past 40 years there have been many calculations of Si\,{\sc ii} transition probabilities and electron impact excitation rates by several authors \citep{1977A&A....58..291N,1991MNRAS.248..827D,1998ADNDT..68..183N,2008ApJS..179..534T},
which have been subsequently employed in modelling codes to predict theoretical line intensities for different types of plasma. However, discrepancies between observation and theory found in 
early studies of the Sun and other late-type stars remain unresolved. For example, \citet{1991MNRAS.253..474D} investigated the 3s$^{2}$3p $^{2}$P$_{J}$--3s3p$^{2}$ $^{4}$P$_{J^{\prime}}$ intercombination multiplet at $\sim$\,2335\,\AA\ in the Skylab spectra of the Sun and International Ultraviolet 
Explorer (IUE) observations of several late-type stars. They noted that, in previous studies, observed line ratios 
involving these lines lay well outside the range of theoretical values, and indeed beyond the high density limit.
This limit depends on the transition probabilities rather than the collisional rates, and hence \citet{1991MNRAS.253..474D} 
suggested that the discrepancies must be due to errors in the former. Although the transition probabilities 
calculated by Dufton et al. 
improved agreement between theory and observation, significant discrepancies remained. 

Similarly, \citet{1991MNRAS.253..123J} undertook an 
observational study of Si\,{\sc ii} emission line intensity ratios in a range of astronomical sources, including the Sun, several late-type stars and the slow nova RR Telescopii. In particular, they analysed a high signal-to-noise,
high resolution spectrum of $\alpha$ Tau, obtained with the Goddard High Resolution Spectrograph (GHRS) on the Hubble Space Telescope (HST). These authors once again found discrepancies between theory and observation for line ratios involving the 2335\,\AA\ multiplet, which they attributed 
in part to blending of the Si\,{\sc ii} lines with Ni\,{\sc ii} transitions. However, they also suggested that the 
discrepancies may be partly due to errors in the adopted electron impact excitation rates, in contrast to the findings of \citet{1991MNRAS.253..474D}.

In the present paper we use recent calculations of atomic data for Si\,{\sc ii} by \citet{2014MNRAS.442..388A} in the modelling code Cloudy \citep{1998PASP..110..761F,2013RMxAA..49..137F} to analyse emission lines in a representative set of ultraviolet spectra of late-type stars and the Sun, to investigate if the discrepancies between theory and observation can be resolved. Specifically, we examine very high quality datasets for three cool stars, $\beta$ Geminorum ($\beta$ Gem), $\alpha$ Centauri A ($\alpha$ Cen A) and B ($\alpha$ Cen B), obtained with the Space Telescope Imaging Spectrograph (STIS) on HST as part of the ASTRAL HST Large Treasury Project (Ayres 2013).  The paper is organised as follows.
Section 2 contains a description of the observations, while Section 3 discusses how we have updated 
the Cloudy database with the most recent Si\,{\sc ii} atomic physics calculations.
In Section 4 we compare the observations with both Cloudy simulations and also those generated with CHIANTI, and summarise our conclusions.

\section{OBSERVATIONAL DATA}
\label{sec:obs}

The main focus of our study are  the three cool stars $\beta$ Gem, $\alpha$ Cen A and B from the ASTRAL HST Large Treasury Project \citep{2013AN....334..105A}. This project is aimed at collecting high signal-to-noise, high spectral resolution ultraviolet spectra for a representative sample of late-type (and subsequently early-type) stars. Observations were obtained with STIS, covering both the FUV (1150--1700\,\AA) and NUV (1600--3100\,\AA)
prime grating settings, at a resolution of R = 100,000, apart from the 1700--2150\,\AA\ region which is at R = 30,000.
For comparison, we note that the resolution of the HST/GHRS spectrum of $\alpha$ Tau analysed by \citet{1991MNRAS.253..123J} was only R = 24,000. 
Our choice of these stars for the \si analysis is due to the fact that they show comparatively narrow unblended emission lines, suitable for the accurate determination and investigation of line fluxes. A detailed description of the ASTRAL data processing and calibration may be found in \citet{2010ApJS..187..149A,2013AN....334..105A}, while further information on the project is available at its website.\footnote{http://casa.colorado.edu/$\sim$ayres/ASTRAL/}

We also study some previously published Si\,{\sc ii} measurements, namely those for the quiet Sun obtained by the HRTS experiment during a sounding rocket flight, which recorded the 1185--1730\,\AA\ solar spectral region on 
photographic emulsion at a resolution of R $\simeq$ 30,000 \citep{1992ApJ...387..726K}. In addition, we re-examine 
the HST/GHRS spectra of the K5\,III star $\alpha$ Tau, originally analysed by \citet{1991MNRAS.253..123J}, which spans the 
2320--2368\,\AA\ region at a resolution of R = 24,000. Further details of the solar and $\alpha$ Tau
observations may be found in \citet{1992ApJ...387..726K} and \citet{1991ApJ...377L..45C}, respectively.

In Table \ref{Table:si2} we list the \si transitions investigated, where we note that vacuum wavelengths are employed
throughout this paper as the HST data are processed using these to avoid a discontinuity at 2000\,\AA\ between vacuum and air wavelengths. After fitting the local continuum with a polynomial, 
we have used a Gaussian profile to model the emission lines. In most cases a Gaussian provides a good fit
to the emission feature, but in some instances (particularly shortward of 1800\,\AA) the Si\,{\sc ii} lines
 show central reversals. These can arise due to chromospheric absorption from the same star or from the intervening 
interstellar medium.

\begin{figure*}
  \centering  
\hbox{
\includegraphics[width=10cm,angle=0]{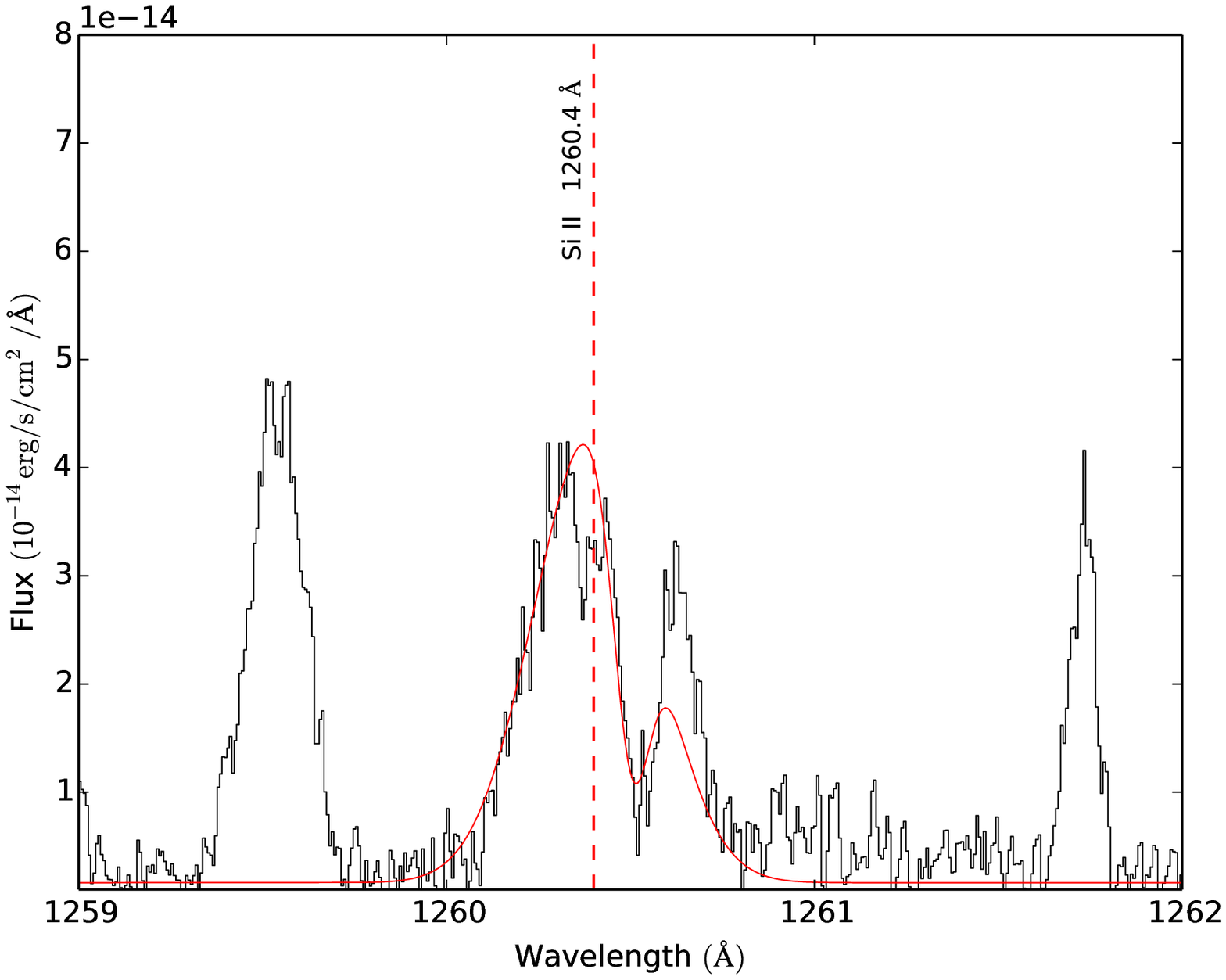} 
\includegraphics[width=10cm,angle=0]{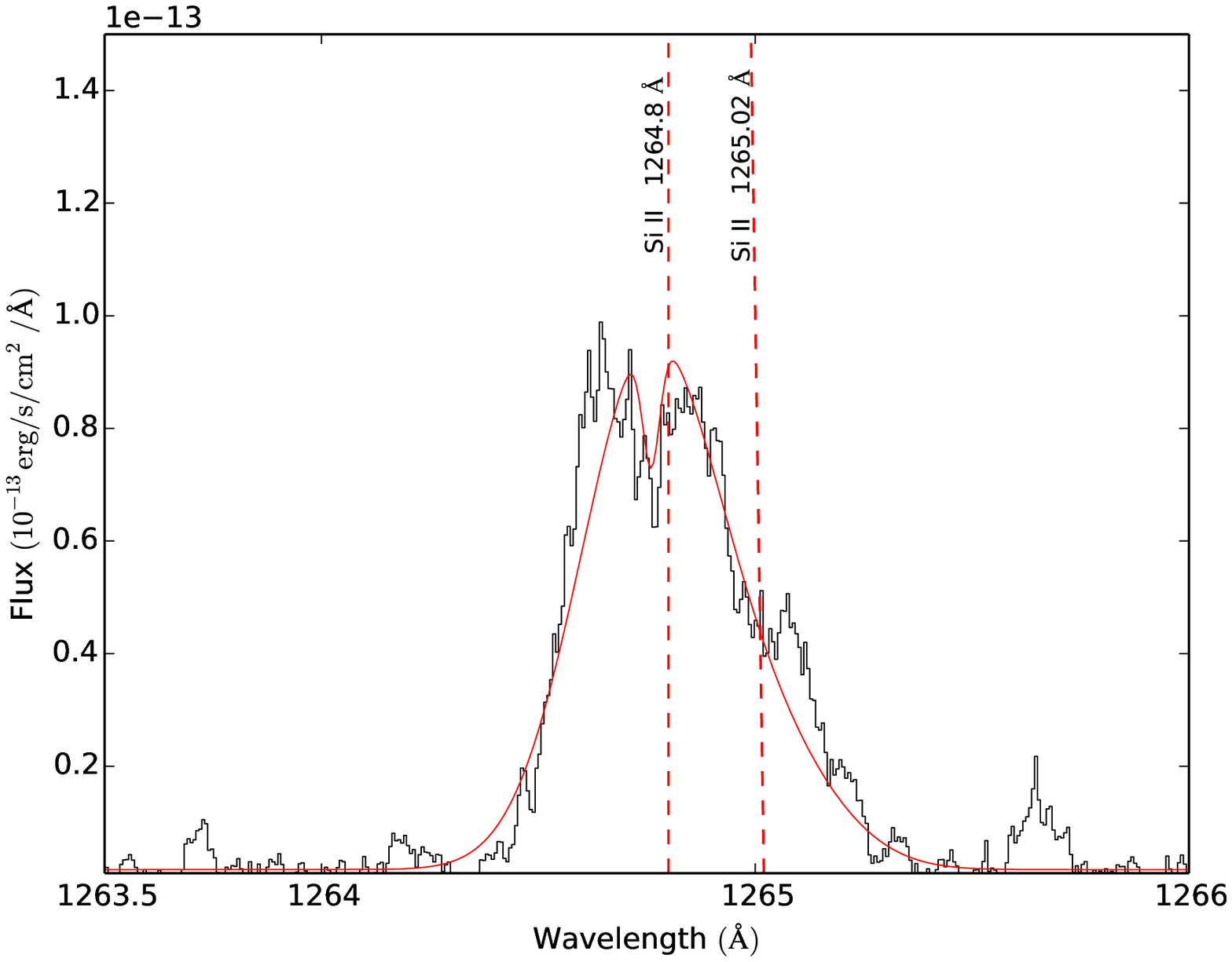} 
}
\caption{Portions of the HST/GHRS spectra of $\beta$\,Gem showing the \si 1260.42 and 1264.73\,\AA\ lines. The best-fit Gaussian profiles to the emission features are shown as solid lines.}\label{fig:1260}
\end{figure*}

\begin{figure*}
  \centering 
\hbox{
\includegraphics[width=10cm,angle=0]{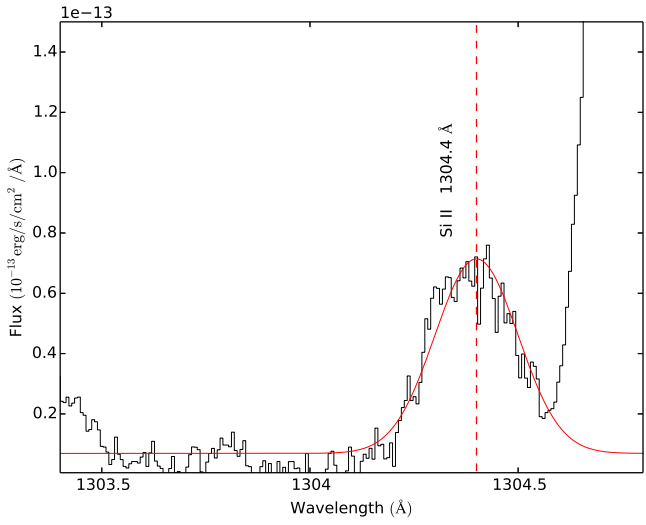} 
\includegraphics[width=10cm,angle=0]{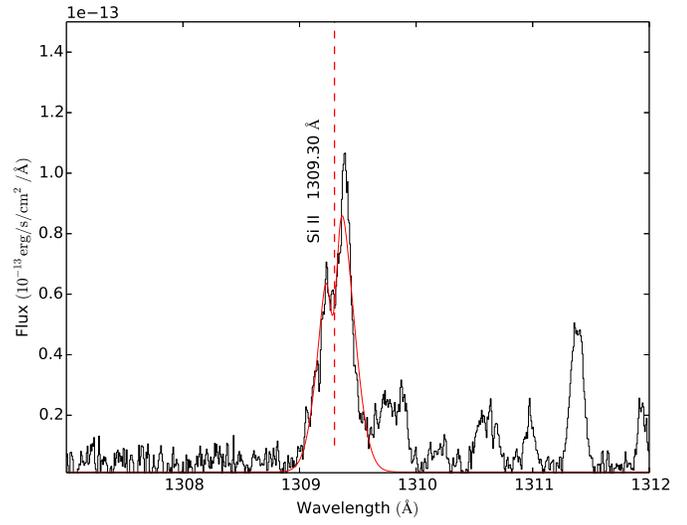} 
}
\caption{Same as Figure 1 except for the \si 1304.37 and 1309.28\,\AA\ lines.}\label{fig:1304}
\end{figure*}

\begin{figure*}
  \centering
\hbox{
\includegraphics[width=10cm,angle=0]{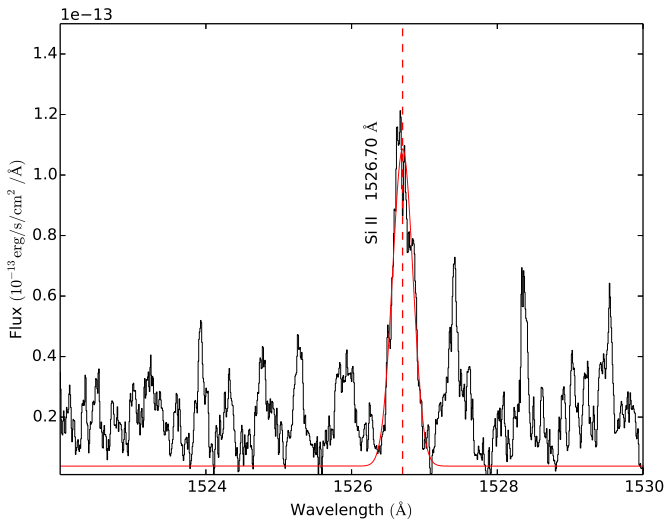} 
\includegraphics[width=10cm,angle=0]{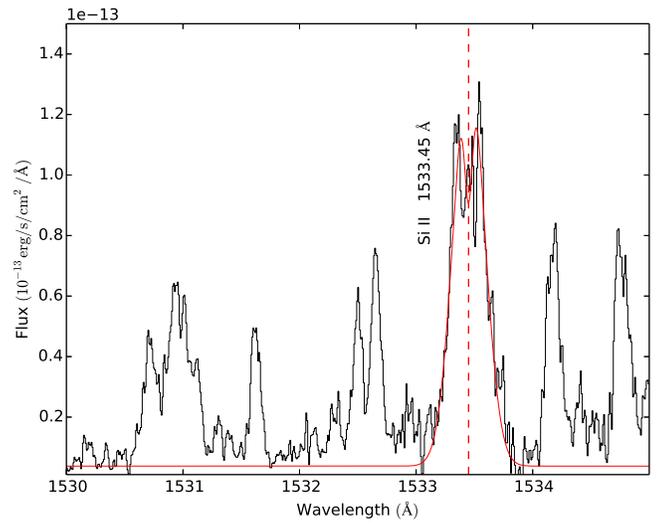} 
}
\caption{Same as Figure 1 except for the \si 1526.70 and 1533.40\,\AA\ lines.}\label{fig:1526}
\end{figure*}

\begin{figure*}
  \centering
\hbox{
\includegraphics[width=10cm,angle=0]{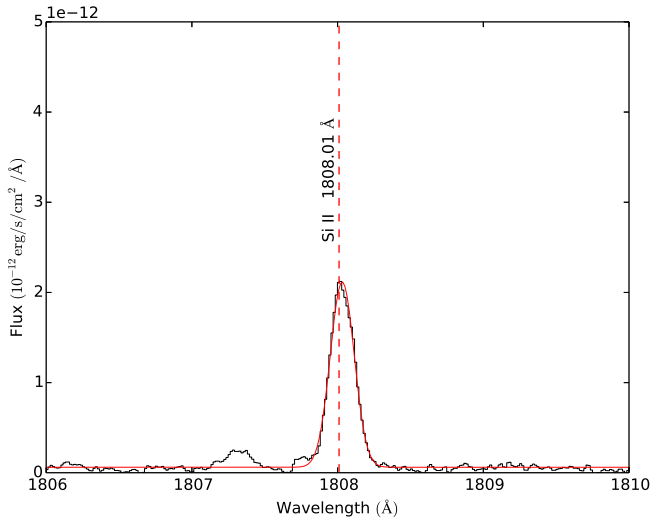} 
\includegraphics[width=10cm,angle=0]{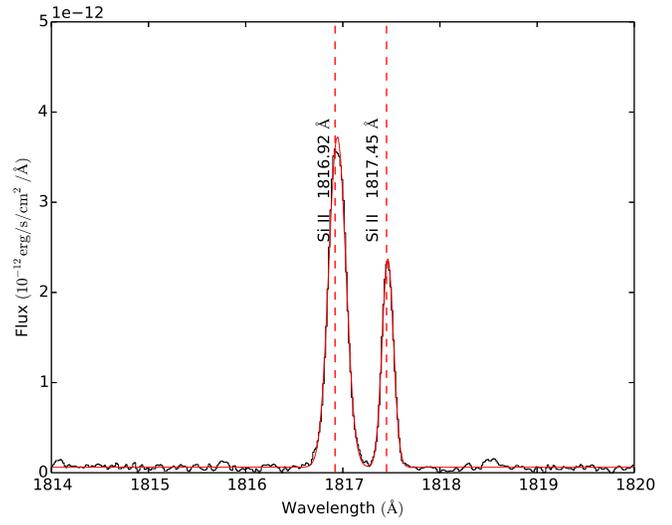} 
}
\caption{Same as Figure 1 except for the \si 1808.01, 1816.93 and 1817.45\,\AA\ lines.}\label{fig:1808}
\end{figure*}

\begin{figure*}
  \centering
\hbox{
\includegraphics[width=10cm,angle=0]{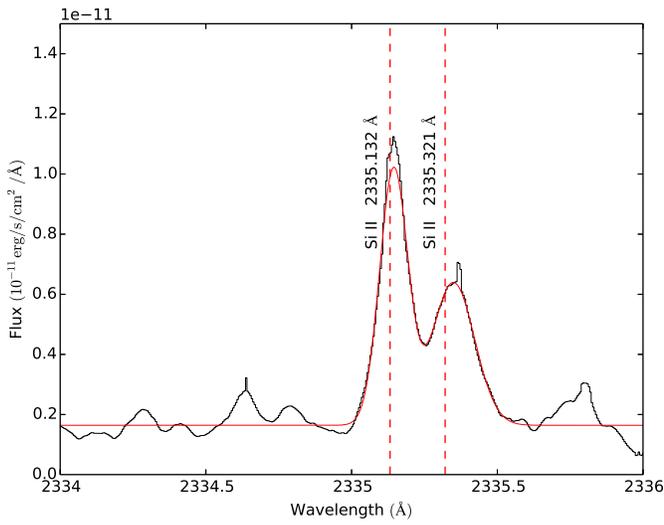} 
\includegraphics[width=10cm,angle=0]{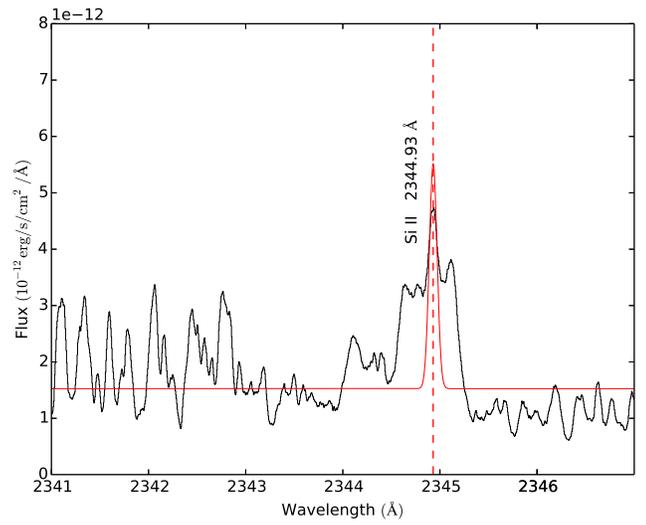} 
}
\caption{Same as Figure 1 except for the \si 2335.12, 2335.32 and 2344.92\,\AA\ lines.}\label{fig:2334}
\end{figure*}

\begin{figure*}
  \centering
\hbox{
\includegraphics[width=10cm,angle=0]{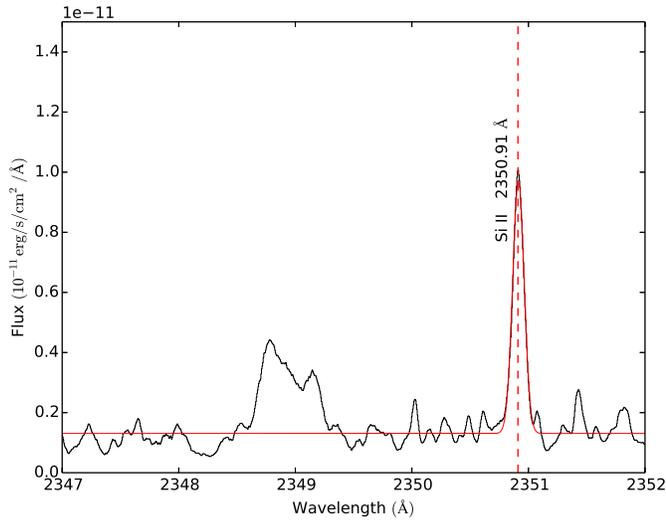} 
}
\caption{Same as Figure 1 except for the \si 2350.89\,\AA\  line.}\label{fig:2350}
\end{figure*}

In Figures \ref{fig:1260}-\ref{fig:2350} we plot portions of the $\beta$ Gem spectrum containing \si emission lines, along with the best-fit Gaussian profiles to the observations, to illustrate the quality of the observational
data. 
As noted above, in some instances there are reversals in the line cores, resulting in an intensity dip. In these cases we fitted the dip in the line profile with an inverted Gaussian as well as the standard one to correct for the reduction in line intensity arising from the dip. The results of the line profile fitting, including measured line wavelengths, widths and intensities, are summarised in Tables \ref{Table:beta-gem} and \ref{Table:alpha}. 
However, for conciseness we only provide wavelengths and widths for $\beta$ Gem (in Table 2).  { The errors in the line
intensities listed in Tables 2 and 3 were derived from a Monte-Carlo simulation of the measurement
process given the assigned photometric error, and are hence lower limits
to the true uncertainties. In some cases, such as for the Si\,{\sc ii} 2344.92\,\AA\ line in $\beta$ Gem,
the error may be dominated by systematic uncertainties, for example how to treat local
blending.}
We note that for 
$\alpha$ Cen A, intensities are only reported for \si features shortward of 1800\,\AA, because the photospheric 
continuum emission overwhelms weak lines (including those of \si) at longer wavelengths.
Below we 
discuss each of the \si multiplets separately. 

\begin{itemize} 

\item 1260\,\AA: (Figure \ref{fig:1260}). There are three emission lines observed in this multiplet (1260.42, 1264.73, 1265.02\,\AA), all of which 
show central reversals and have complex profiles. The intensity of 1260.42\,\AA\ has been measured 
by numerical integration of the blue half of the line profile (which is least affected by central reversal), and then doubled to obtain the total line flux. In Figure \ref{fig:1260} we plot the 1260.42\,\AA\ feature fitted 
with two Gaussian profiles, where the negative Gaussian models the central reversal. The lines at 
1264.73 and 1265.02\,\AA\ are blended, and we have used two Gaussian profiles to model these, with the central reversal in the former fitted using a negative Gaussian.

\item 1304\,\AA: (Figure \ref{fig:1304}). Both the lines in this multiplet (1304.37, 1309.28\,\AA) are slightly broader than the instrument resolution. 
There is a very strong O\,{\sc i} line at 1304.87\,\AA. The \si line at 1309.28\,\AA\ shows prominent central reversal which was modelled using a negative Gaussian profile.  

\item 1526\,\AA: (Figure \ref{fig:1526}). Both the lines in this multiplet (1526.70, 1533.40\,\AA) are broader than the instrument resolution. 

\item 1808\,\AA: (Figure \ref{fig:1808}). The three lines observed in this multiplet (1808.01, 1816.93, 1817.45\,\AA) are narrow and unblended. 

\item 2335\,\AA: (Figures \ref{fig:2334} and \ref{fig:2350}). There are four observed lines in this multiplet (2335.12, 2335.32, 2344.92, 2350.89\,\AA), with the 
2329.23\,\AA\ transition not detected due to its very small A-value and hence predicted intensity. 
The \si 2335.32\,\AA\ line is blended with \Ni 2335.30\,\AA. However, \citet{1988ApJS...68..345C} note that the \Ni line is in fact radiatively pumped by the \si transition, as fluorescence emission is detected in another \Ni feature at 2416.87\,\AA. Hence, as noted by \citet{1991MNRAS.253..123J}, 
the photons observed in the \Ni 2416.87\,\AA\ line must actually be \si photons and the intensities of the \Ni and \si transitions need to be added. Furthermore, \citet{1991MNRAS.253..123J}  pointed out that the pumped \Ni level also gives rise to additional lines at 2084.31, 2125.79 and 2161.89\,\AA. Unfortunately, the spectra analysed by \citet{1991MNRAS.253..123J} could not observe these three lines as they were outside the wavelength range of the HST/GHRS observations, and the IUE detectors were insensitive at these wavelengths. These authors hence had to calculate their contribution to the \si 2335.32\,\AA\ line flux. However, the STIS spectra for our cool star sample
do contain these \Ni features, along with the 2416.87\,\AA\ line, as may be seen in Figures 7--10 for $\beta$ Gem. Their measured wavelengths, widths and intensities are summarised in Table 2 for $\beta$ Gem, while line intensities 
are listed for $\alpha$ Cen B in Table 3.
We note that Judge et al. calculate the ratio of the total intensity of the three additional Ni\,{\sc ii} lines to that of the 2416.87\,\AA\ feature to be 1.1, while from our STIS observations we find this ratio to be 1.1 for 
$\beta$ Gem and 0.48 for $\alpha$ Cen B. 

\end{itemize}

The resultant \si line intensity ratios from the $\beta$ Gem, $\alpha$ Cen A and B observations are listed in Table \ref{Table:ratio1}, where we note that the \Ni line fluxes have been added to that for \si 2335.32\,\AA\ in the determination of the 2335.32/2350.89 ratio. In Table \ref{Table:ratio2} we provide some additional measurements of \si line ratios 
for $\alpha$\,Tau from the HST/GHRS observations of \citet{1991MNRAS.253..123J}, and HRTS quiet Sun data from \citet{1992ApJ...387..726K}.

\section{THEORETICAL LINE RATIOS}

We have undertaken several \si line intensity ratio calculations 
using the Cloudy modelling code with two sets of atomic data, to investigate the importance of
adopting different atomic physics parameters on the theoretical results. All three Cloudy models generated
consist of the energetically-lowest 148 fine-structure levels, with the energies taken from the NIST 
 database\footnote{http://www.nist.gov/pml/data/asd.cfm} as at 2015 August 8. 
 The first atomic model, denoted CLOUDY1, 
 employs the experimental A-values of \citet{1993ApJ...415L..59C} for the 
 3s$^{2}$3p $^{2}$P$_{J}$--3s3p$^{2}$ $^{4}$P$_{J^{\prime}}$ 
intercombination lines, and the calculations of \citet{1998ADNDT..68..183N} for the remaining transitions, plus the new Si\,{\sc ii} 
electron impact effective collision strengths (ECS) of \citet{2014MNRAS.442..388A}, which are not too different from the ECS data of \citet{1994ADNDT..57..273D} adopted in earlier versions of Cloudy. In the second atomic model, CLOUDY2, 
the A-values for transitions among the lowest 56 fine-structure levels are 
replaced by the calculations of \citet{2014MNRAS.442..388A}, which in many instances show large discrepancies with the results of \citet{1993ApJ...415L..59C} and \citet{1998ADNDT..68..183N}. 
We have also calculated Si\,{\sc ii} line ratios 
with the CHIANTI atomic database version 7.1.4 \citep{1997A&AS..125..149D,2013ApJ...763...86L}, which employs the A-value calculations of Tayal (2007), Nahar (1998) and Nussbaumer (1977), 
plus ECS from \citet{2008ApJS..179..534T}. The Calamai et al., Tayal (2008), Nahar and Aggarwal \& Keenan
atomic data for the Si\,{\sc ii} lines discussed in the present paper are summarised in Table 1 for comparison purposes. The CLOUDY1, CLOUDY2 and CHIANTI models were calculated for optically thin plasma conditions, and hence for completeness we have produced an additional Cloudy model, CLOUDY3, which uses the same atomic data as CLOUDY2 but calculates line ratios for an optically thick plasma, specifically a uniform slab with 
a total hydrogen column density of 10$^{23}$ cm$^{-2}$.

\section{RESULTS AND DISCUSSION}

In Table \ref{Table:ratio1} we summarise the observed \si emission line intensity ratios for 
$\beta$ Gem, $\alpha$ Cen A and B, along with theoretical optically thin values from the CHIANTI, CLOUDY1 and CLOUDY2 models calculated at an electron temperature T$_{e}$ = 7000\,K and density N$_{e}$ = 10$^{9.5}$\,cm$^{-3}$, typical of the physical conditions found in the line-forming regions of chromospheres in late-type giant stars \citep{1986MNRAS.221..119J,1986MNRAS.223..239J}. For the CLOUDY2 model we also 
list theoretical ratios at N$_{e}$ = 10$^{8.5}$ and 10$^{10.5}$\,cm$^{-3}$, to show their sensitivity to the adopted electron density. (We note that the ratios do not vary significantly with temperature). In addition, results are presented 
for an optically thick plasma (CLOUDY3 model) to allow the effect of opacity on the results to be assessed, in particular for ratios involving strong allowed lines.

Similarly, in Table \ref{Table:ratio2} we list observed line ratios for $\alpha$\,Tau from the HST/GHRS observations of \citet{1991MNRAS.253..123J}, plus HRTS quiet Sun data from \citet{1992ApJ...387..726K}. 
In addition we provide CHIANTI, CLOUDY1, CLOUDY2 and CLOUDY3 model results, calculated at the same plasma parameters for $\alpha$\,Tau as for the other cool stars, i.e. 
T$_{e}$ = 7000\,K and N$_{e}$ = 10$^{9.5}$\,cm$^{-3}$, as the former is also a late-type giant. For the quiet Sun we have generated theoretical ratios at T$_{e}$ = 7000\,K and N$_{e}$ = 10$^{11}$\,cm$^{-3}$ (Dufton \& Kingston 1983),
but note that the ratios considered in Table \ref{Table:ratio2} are insensitive to the adopted plasma parameters. 

The first point to note from an inspection of Tables \ref{Table:ratio1} and \ref{Table:ratio2} is that the theoretical line ratios from the CHIANTI and 
CLOUDY1 models do not in general differ
 significantly, with discrepancies of typically only 11\%. As the major difference between these models is the adoption of the new \citet{2014MNRAS.442..388A} ECS in the CLOUDY1 calculations, we can state that use of the \citet{2014MNRAS.442..388A} data does not significantly change the resultant Si\,{\sc ii} line ratios, at least for those involving the 
 ultraviolet transitions considered in the present paper. 
This is perhaps not surprising, as the \citet{2014MNRAS.442..388A} ECS are in generally good agreement with previous results, as shown in Table \ref{Table:si2} for the \citet{2008ApJS..179..534T} data.

{ However, the situation is very different when the new A-value calculations of \citet{2014MNRAS.442..388A} are incorporated in plasma models, with the CLOUDY2 line ratios showing large discrepancies in some instances with the CHIANTI and CLOUDY1 results. In particular, the 1808.01/1817.45 ratio (3s$^2$3p $^2$P--3s3p$^2$ $^2$D transitions) and those involving the intercombination multiplet lines around 
$\sim$2335\,\AA\ (3s$^2$3p $^2$P--3s3p$^2$ $^4$P) in the CLOUDY2 calculations are up to a factor of 7 smaller than the CHIANTI and CLOUDY1 data. This is due to the A-values of \citet{2014MNRAS.442..388A} for the corresponding transitions similarly being much lower than those of \citet{1998ADNDT..68..183N}, by more than an order of magnitude in some cases (see Table 1). Aggarwal \& Keenan discuss their calculations in great detail, and compare these with previous results. They estimate their A-values to be generally accurate to $\pm$~20\%, although this only applies to strong allowed transitions. For weaker ones, such as the 1808 and 2335\,\AA\ multiplet lines, the errors are likely to be larger and are very difficult to quantify, as discussed by Aggarwal \& Keenan. It is unclear why there are large  discrepancies between the Aggarwal \& Keenan  data and the theoretical work of \citet{1998ADNDT..68..183N}, and more importantly between the former and the experimental values of \citet{1993ApJ...415L..59C} for the 3s$^2$3p $^2$P$_J$--3s3p$^2$ $^4$P$_{J^\prime}$  intercombination lines. However, we point out that the A-values for the 
3s$^2$3p $^2$P--3s3p$^2$ $^2$D lines are up to three orders of magnitude smaller than those for 
the 3s$^2$3p $^2$P--3s3p$^2$ $^2$P allowed transitions (see Table 3 of Aggarwal \& Keenan). 
As noted by the referee, since a large part of the matrix element for the 3s$^2$3p $^2$P$_J$--3s3p$^2$ $^4$P$_{J^\prime}$  intercombination transitions arises from fine-structure interactions between the 3s3p$^2$ $^4$P$_J$ and $^2$D$_J$  levels, these becomes very sensitive because of the smaller size of the latter. Consequently, the A-values for the 1808 and 2335\,\AA\ multiplets are connected and very difficult to calculate. We hence strongly recommend further work on the determination of A-values for Si II, and in particular new measurements for the intercombination multiplet.}

Given the discrepancies between the \citet{2014MNRAS.442..388A} results and previous theoretical and experimental data for A-values, it would be understandable to assume that the former must be in error. However, an inspection of Tables \ref{Table:ratio1} and \ref{Table:ratio2} reveal that the CLOUDY2 line ratio calculations, which include the \citet{2014MNRAS.442..388A} A-values, are in mostly better agreement with the observations than the CHIANTI and CLOUDY1 results. In particular, the long-standing 
discrepancies between theory and observation for the intercombination line ratios \citep[see, for example][and references therein]{1991MNRAS.253..123J,1991MNRAS.253..474D} are removed. Similarly, there is a significant improvement in the agreement between experimental values of the 1808.01/1817.45 line ratio and the theoretical results. This may be a coincidence, but we believe it is unlikely that the A-values for the components of the
3s$^{2}$3p $^{2}$P$_{J}$--3s3p$^{2}$ $^{4}$P$_{J^{\prime}}$ 
intercombination multiplet and the
3s$^{2}$3p $^{2}$P$_{J}$--3s3p$^{2}$ $^{2}$D$_{J^{\prime}}$ 
allowed lines would both be in error by the precise amounts required to provide good agreement between
theory and observation, as they are different types of transition, {albeit interconnected as noted above}. 
In addition, the \citet{2014MNRAS.442..388A} A-values for other Si\,{\sc ii} transitions are in agreement with previous work, with the resultant CLOUDY2 ratios similar to those generated by CHIANTI and CLOUDY1, which in turn generally agree 
with the observations. If the \citet{2014MNRAS.442..388A} data were in error, 
we might expect other transitions to be similarly affected. 
On the other hand, the \citet{2014MNRAS.442..388A} results imply that not only are previous calculations in error, but so are the measurements for the Si\,{\sc ii} intercombination lines by \citet{1993ApJ...415L..59C}, by up to a factor of 8, which seems unlikely. 
As noted above, new experimental determinations of the Si\,{\sc ii} A-values are urgently required to investigate if the previous data are in error. If not, then the discrepancies between theory and observation for the intercombination lines remain, and will need further investigation. 

In the cases of the 2329.23/2350.89 and 2335.32/2350.89 ratios in Table \ref{Table:ratio2} for $\alpha$\,Tau, 
the CLOUDY2 values are in poorer agreement with observations than the CHIANTI and CLOUDY1 results.
However, as noted previously the 2335.32\,\AA\ line intensity needs to be added to those for Ni\,{\sc ii} features which are pumped by Si\,{\sc ii} emission. Although we have identified and measured line intensities for all four 
Ni\,{\sc ii} transitions in the spectrum of $\beta$ Gem and for three of the lines in $\alpha$ Cen B (as
2084.31\,\AA\ was not detected in this star, \citet{1991MNRAS.253..123J} could only determine the intensity for Ni\,{\sc ii} 2416.87\,\AA\ and had to calculate fluxes for the others. Hence their estimate for the 
2335.32\,\AA\ line intensity may not be reliable.
Similarly, the 2329.23\,\AA\ feature is weak in the HST/GHRS spectrum of $\alpha$\,Tau (it was not detected in the 
higher quality HST/STIS data for the other stars) and hence may also not be well determined.

We note that although the 1260.42/(1264.73+1265.02) ratio for the quiet Sun in Table \ref{Table:ratio2} agrees with all the theoretical optically thin values, the cool stars measurements for 1260.42/1264.73 in Table \ref{Table:ratio1} are somewhat smaller than predicted. This cannot be due to 
optical depth effects, as the CLOUDY3 calculation for an optically thick plasma shows an even larger discrepancy with
observation. The most likely explanation for the disagreement would be blending in the 1264.73\,\AA\ line, but a synthetic spectrum generated with CHIANTI reveals no likely candidate. For the other ratios involving allowed lines, the CLOUDY3 results indicate that at least some of the transitions must be subject to significant opacity, as the observations are in better agreement with these than with the optically thin CHIANTI, CLOUDY1 and CLOUDY2 
theoretical ratios, one example being 1526.70/1533.40 in the three cool stars plus the quiet Sun.

To summarise, \si line intensity ratios measured from the ultraviolet spectra of cool stars and the Sun are found to be in generally good agreement with theoretical results generated with the Cloudy modelling code 
which include the radiative rate calculations of \citet{2014MNRAS.442..388A}. In particular, adopting their A-values removes discrepancies previously found between theory and observation for ratios involving the 
3s$^{2}$3p $^{2}$P$_{J}$--3s3p$^{2}$ $^{4}$P$_{J^{\prime}}$ 
intercombination transitions at $\sim$\,2335\,\AA. However, these A-values are significantly different 
(by up to a factor of 8) from both previous calculations and experimental determinations. New measurements of the
intercombination line A-values are hence urgently required to investigate if the existing experimental data are wrong. If not, then the good agreement found between theory and observation in the present paper is simply
a coincidence.

\section*{ACKNOWLEDGEMENTS}

Based on observations made with the NASA/ESA Hubble Space Telescope, obtained from the Mikulsky Archive at Space Telescope Science Institute, operated by the Association of Universities for Research in Astronomy, Inc., under NASA contract NAS 5-26555. Support for ASTRAL is provided by grants HST-GO-12278.01-A and HST-GO-13346.01-A from STScI. The project has made use of public databases hosted by SIMBAD, maintained by CDS, Strasbourg, France. SL, CAR and FPK are grateful to STFC for financial support via grant ST/L000709/1. GJF acknowledges financial support from the Leverhulme Trust via Visiting Professorship grant VP1-2012-025. CHIANTI is a collaborative project involving George Mason University, the University of Michigan (USA) and the University of Cambridge (UK).



\begin{table*}
\centering
\begin{minipage}{140mm}
\caption{\si emission lines studied in the present work.$^{1}$} 
  \begin{tabular}{llccccccccccccc}
   \hline 
  Lower level			& Upper level 					& Wavelength (\AA) 	& A$_{ij}^{\rm expt}$& A$_{ij}^{\rm N}$	
  & A$_{ij}^{\rm AK}$	& ECS$_{ij}^{\rm T}$	& ECS$_{ij}^{\rm AK}$	
  \\
($i$)				& ($j$)						& (NIST) 			& (s$^{-1}$)	  	&(s$^{-1}$)			
& (s$^{-1}$)		& (T$_{e}$ = 7000 K)		& (T$_{e}$ = 7000 K)	
\\
 \hline                                                                                                                                                                                                   
3s$^2$3p $^2$P$_{1/2}$	& 3s$^2$3d $^2$D$_{3/2}$ 		&  1260.42 	& \ldots				& 2.60E+9 			& 2.01E+9            & 3.42				& 2.31
\\                                         
3s$^2$3p $^2$P$_{3/2}$	& 3s$^2$3d $^2$D$_{5/2} $	&  1264.73 	& \ldots				& 3.04E+9 			& 2.31E+9             & 6.76 				& 4.71
\\ 
3s$^2$3p $^2$P$_{3/2}$	& 3s$^2$3d $^2$D$_{3/2} $	&  1265.02 	& \ldots				& 4.63E+8 			& 5.23E+8             & 1.85 				& 1.37
\\
3s$^2$3p $^2$P$_{1/2}$	& 3s3p$^2$ $^2$S$_{1/2}$	&  1304.37 	& \ldots		& 3.64E+8 			& 3.60E+8             & 1.00 				& 0.89
\\
3s$^2$3p $^2$P$_{3/2}$	& 3s3p$^2$ $^2$S$_{1/2}$  	&  1309.28 	& \ldots		& 6.23E+8 			& 6.60E+8             & 1.98 				& 1.79
\\
3s$^2$3p $^2$P$_{1/2}$	& 3s$^2$4s $^2$S$_{1/2}$ 	&  1526.70  & \ldots	& 3.81E+8 & 3.90E+8            & 1.06 				& 1.21
\\
3s$^2$3p $^2$P$_{3/2}$	& 3s$^2$4s $^2$S$_{1/2}$   	&  1533.40	& \ldots	& 7.52E+8 			& 7.90E+8             & 2.13 				& 2.43
\\
3s$^2$3p $^2$P$_{1/2}$	& 3s3p$^2$ $^2$D$_{3/2}$	&  1808.01 	& \ldots	& 2.54E+6 			& 1.00E+5             & 2.77 				& 1.91
\\
3s$^2$3p $^2$P$_{3/2}$	& 3s3p$^2$ $^2$D$_{5/2}$	&  1816.93		& \ldots		& 2.65E+6 			& 2.00E+5             & 7.46 				& 5.25
\\ 
3s$^2$3p $^2$P$_{3/2}$	& 3s3p$^2$ $^2$D$_{3/2}$ 	&  1817.45 	& \ldots				& 3.23E+5 	& 5.30E+4             & 4.17 		& 3.08
\\
3s$^2$3p $^2$P$_{1/2}$   & 3s3p$^2$ $^4$P$_{3/2}$ 	&  2329.23     	& 10 $\pm$ 50			& 23.5				& 11.1                 & 0.80				& 0.75
\\
3s$^2$3p $^2$P$_{1/2}$   & 3s3p$^2$ $^4$P$_{1/2}$  	&  2335.12  	& 5200$ \pm$ 19	& 5510	& 2296              & 0.51				& 0.47
\\
3s$^2$3p $^2$P$_{3/2}$   & 3s3p$^2$ $^4$P$_{5/2}$  		&  2335.32       	& 2460 $\pm$ 8	& 2440		& 397                & 2.38				& 2.07
\\
3s$^2$3p $^2$P$_{3/2}$   & 3s3p$^2$ $^4$P$_{3/2}$  	&  2344.92        	& 1220 $\pm$ 10			& 1310				& 157                 & 1.05				& 1.04
\\ 
3s$^2$3p $^2$P$_{3/2}$   & 3s3p$^2$ $^4$P$_{1/2}$  	&  2350.89       	& 4410 $\pm$ 21	& 4700	& 3078                 & 0.31				& 0.40
\\
 \hline 
\end{tabular} \label{Table:si2} 

$^{1}$A$^{\rm expt}$ are the experimental A-values from \citet{1993ApJ...415L..59C}; A$^{\rm N}$ and ECS$^{\rm T}$ the A-values and effective collision strengths from \citet{1998ADNDT..68..183N} and \citet{2008ApJS..179..534T}, respectively;
A$^{\rm AK}$ the wavelength-corrected transition probabilities and ECS$^{\rm AK}$ the effective collision strengths from \citet{2014MNRAS.442..388A}.
\end{minipage}
\end{table*}


\begin{table*}
\centering
\begin{minipage}{140mm}
\caption{Wavelengths, widths and intensities for the Si\,{\sc ii} and Ni\,{\sc ii} emission lines 
in the HST/STIS spectrum of $\beta$\,Gem.} 
  \begin{tabular}{lccc}
   \hline 
Experimental wavelength (\AA)	& Line centroid (\AA)  	&  Line width (\AA)	& Line flux 
\\ 
(in vacuum)		  	& (observed)		&  (observed)		&  (10$^{-14}$\,erg\,cm$^{-2}$\,s$^{-1}$) 
\\ 
\hline 
1260.42 			& 1260.35 $\pm$ 0.01  	& 0.136 $\pm$ 0.007	& 1.75 $\pm$ 0.08
\\  
1264.73				& 1264.80 $\pm$ 0.01   	& 0.188 $\pm$ 0.003 	& 3.97 $\pm$ 0.09
\\
1265.02     		& 1265.20  $\pm$ 0.01	& 0.160 $\pm$ 0.003	& 1.57 $\pm$ 0.06
\\
1304.37	 			& 1304.40 $\pm$ 0.01 	& 0.099 $\pm$ 0.001  	& 1.85 $\pm$ 0.08
\\ 
1309.28	 			& 1309.34 $\pm$ 0.02 	& 0.116 $\pm$ 0.003	& 2.59 $\pm$ { 0.06}
\\ 
1526.70	 			& 1526.70 $\pm$ 0.01  	& 0.113 $\pm$ 0.003 	& 3.60 $\pm$ 0.09
\\                                     
1533.40				& 1533.45 $\pm$ 0.01   	& 0.131 $\pm$ 0.006	& 4.15 $\pm$ 0.15
\\ 
1808.01				& 1808.03 $\pm$ 0.01   	& 0.083 $\pm$ 0.003 	& 43.6 $\pm$ 0.1
\\ 
1816.93		 		& 1816.94 $\pm$ 0.01 	& 0.084 $\pm$ 0.001 	& 82.7 $\pm$ 0.1
\\
1817.45				& 1817.46 $\pm$ 0.01   	& 0.059 $\pm$ 0.001 	& 34.8 $\pm$ 0.1
\\
2329.23$^1$			& \ldots 	  	& \ldots		& \ldots
\\
2335.12				& 2335.12 $\pm$ 0.01   	& 0.048 $\pm$ 0.001	& 124 $\pm$ 1
\\
2335.32$^2$			& 2335.32 $\pm$ 0.01 	& 0.070 $\pm$ 0.001	& 78.6 $\pm$ { 0.3}
\\                                       
2344.92				& 2344.93 $\pm$ 0.01 	& 0.040 $\pm$ 0.001	& 61.1 $\pm$ { 0.4} 
\\ 
2350.89				& 2350.91 $\pm$ 0.01 	& 0.050 $\pm$ 0.001	& 104 $\pm$ 1 
\\
2084.31	(\Ni)$^3$		& 2084.30 $\pm$ 0.01 	& 0.055 $\pm$ 0.002 	& 2.52 $\pm$ 0.17
\\  
2125.79 (\Ni)$^3$		& 2125.79 $\pm$ 0.01 	& 0.048 $\pm$ 0.001 	& 11.3 $\pm$ 0.2
\\  
2161.89	(\Ni)$^3$		& 2161.88 $\pm$ 0.01 	& 0.053 $\pm$ 0.001 	& 13.4 $\pm$ 0.2
\\  
2416.87	(\Ni)$^3$		& 2416.90 $\pm$ 0.01 	& 0.050 $\pm$ 0.001 	& 24.5 $\pm$ 0.3
\\
 \hline 
\end{tabular} \label{Table:beta-gem}

$^{1}$Line not detected in the $\beta$\,Gem spectrum.
\\
$^{2}$Blended with \Ni 2335.20\,\AA.
\\
$^{3}$Line of \Ni which is radiatively pumped by \si 2335.32\,\AA. See Section \ref{sec:obs} for 
details. 
\end{minipage}
\end{table*}


\begin{table*}
\centering
\begin{minipage}{140mm}
\caption{Intensities for the Si\,{\sc ii} and Ni\,{\sc ii} emission lines 
in the HST/STIS spectra of $\alpha$ Cen A and $\alpha$ Cen B.} 
  \begin{tabular}{lccc}
   \hline 
Experimental wavelength (\AA)		& Line flux ($\alpha$ Cen A)               & Line flux ($\alpha$ Cen B)          
\\                                                                                                 
(in vacuum)		  	     	& (10$^{-14}\funit$)   		     	    &  (10$^{-14}\funit$)
\\                                                                                                 
\hline                                                                                             
1260.42 				& 9.73 $\pm$ 0.10                          	& 6.95 $\pm$ 0.10
\\                                                                                                 
1264.73				 	& 22.9 $\pm$ 0.10                            & 22.7 $\pm$ 0.3
\\ 
1265.02				 	& 9.11 $\pm$ 0.08                            & 6.32 $\pm$ 0.05
\\                                                                                                  
1304.37	 			  	& 7.95 $\pm$ 0.07                           & 5.00 $\pm$ 0.06
\\                                                                                                 
1309.28	 				& 14.1 $\pm$ 0.1                           &  8.09 $\pm$ 0.06
\\                                                                                                 
1526.70	 			 	& 32.6 $\pm$ 0.2                            & 17.9 $\pm$ 0.1
\\                                                                                                 
1533.40					& 35.4 $\pm$ 0.2                            & 19.8 $\pm$ 0.2
\\                                                                                                 
1808.01				 	& 322 $\pm$ 2                               & 179 $\pm$ 1
\\                                                                                                 
1816.93		 		 	& 508 $\pm$ 2                              &  306 $\pm$ 1
\\                                                                                                 
1817.45				 	& 243 $\pm$ 2                               & 95.6 $\pm$ 0.4
\\                                                                                                 
2329.23$^1$			    	 & \ldots                                    & \ldots
\\                                                                                                 
2335.12					&   \ldots                             & 127 $\pm$ 1
\\                                                                                                 
2335.32$^2$				&   \ldots                               & 74.9 $\pm$ 0.3
\\                                                                                                 
2344.92					&    \ldots                               & 135 $\pm$ 1 
\\                                                                                                 
2350.89					&    \ldots                            & 108 $\pm$ 1
\\                                                                                                 
2084.31	(\Ni)$^3$	 		& \ldots	                          & \ldots
\\                                                                                                 
2125.79 (\Ni)$^3$	 		&  \ldots                           & 10.3 $\pm$ 0.2
\\                                                                                                 
2161.89	(\Ni)$^3$	 		&   \ldots                             & 13.7$\pm$ 0.2
\\                                                                                                 
2416.87	(\Ni)$^3$	 		& \ldots                             & 49.1 $\pm$ 0.7
\\
 \hline 
\end{tabular} \label{Table:alpha}

$^{1}$Line not detected in the $\alpha$ Cen A and $\alpha$ Cen B spectra.
\\
$^{2}$Blended with \Ni 2335.20\,\AA.
\\
$^{3}$Line of \Ni which is radiatively pumped by \si 2335.32\,\AA. See Section \ref{sec:obs} for 
details. 
\end{minipage}
\end{table*}


\begin{table*}
\centering
\begin{minipage}{140mm}
\caption{Comparison of observed and theoretical Si\,{\sc ii} line intensity ratios for $\beta$\,Gem,
$\alpha$ Cen A and B.} 
  \begin{tabular}{lccccccccc}
   \hline 
Line ratio		&  Observed 		&  Observed 		&  Observed 		& CHIANTI$^{1}$	& CLOUDY1$^{1}$	& CLOUDY2$^{1,2}$	& CLOUDY3$^{3}$
\\			& $\beta$~Gem 		&  $\alpha$~Cen~A	& $\alpha$~Cen~B \\                                                                                          
\hline \\                                                                                         
1260.42/1264.73		& 0.44 $\pm$ 0.02 	& 0.42 $\pm$ 0.01 	& 0.30 $\pm$ 0.01 	& 0.56			& 0.56			 & 0.55	(0.55, 0.55)		& 0.97	
\\
1264.73/1265.02   & { 2.52} $\pm$ 0.12   & 2.51 $\pm$ 0.02    &  3.59 $\pm$ 0.05   &  9.55    &    9.69    &  9.41 (9.37, 9.41)   & 0.96
\\                                                                                                 
1260.42/1304.37		& 0.94 $\pm$ 0.06 	& 1.22 $\pm$ 0.02 	& 1.39 $\pm$ 0.03 	& 2.31			& 1.83			& 1.77 (1.91, 1.56)		& 0.22	
\\	                                                                                        
1304.37/1309.28		&  0.72  $\pm$ 0.09	& 0.56 $\pm$ 0.01	& 0.62 $\pm$ 0.01	& 0.58			& 0.59			& 0.55	(0.55, 0.55)		& 0.95	
\\                                                                                              
1526.70/1533.40		& 0.87 $\pm$ 0.04	& 0.92 $\pm$ 0.01	& 0.90 $\pm$ 0.01	& 0.51			& 0.51			& 0.51 (0.51, 0.51)		& 0.96	
\\                                                                                              
1526.70/1808.01		& 0.082 $\pm$ 0.010	& 0.10 $\pm$ 0.01	& 0.10 $\pm$ 0.01	& 0.026			& 0.039			& 0.050 (0.051, 0.042)		& 0.32	
\\                                                                                              
1808.01/1817.45		& 1.25 $\pm$ 0.01	& 1.32 $\pm$ 0.01	& 1.87 $\pm$ 0.01	& 7.47			& 7.90			& 1.93 (1.93, 1.93)		& 0.87	
\\                                                                                              
1816.93/1808.01		& 1.89 $\pm$ 0.01	& 1.58 $\pm$ 0.01	& 1.71 $\pm$ 0.01	& 1.69			& 1.68			& 2.31	(2.27, 2.35)		& 1.03	
\\                                                                                              
2335.12/2350.89		& 1.19 $\pm$ 0.02	& \ldots		& 1.18 $\pm$ 0.01	& 1.20			& 1.18			& 0.75	(0.75, 0.75)		& 0.96	
\\                                                                                              
2335.32/2350.89		& 1.25 $\pm$ 0.01$^{4}$	& \ldots		& 1.37 $\pm$ 0.01	& 4.14			& 4.41			& 1.42	(3.71, 0.50) & 1.12	
\\                                                                                              
2344.92/2350.89		& 0.58 $\pm$ 0.01	& \ldots		& 1.25 $\pm$ 0.02	& 2.04			& 2.92			& 0.41	(1.71, 0.13)		& 1.00	
\\ 
\hline 
\end{tabular} \label{Table:ratio1}

$^{1}$Optically thin line ratios calculated at T$_{e}$ = 7000\,K, N$_{e}$ = 10$^{9.5}$\,cm$^{-3}$.
\\
$^{2}$Values in brackets are optically thin line ratios calculated at T$_{e}$ = 7000\,K, N$_{e}$ = 10$^{8.5}$ and
10$^{10.5}$\,cm$^{-3}$, respectively.
\\
$^{3}$Optically thick calculations at T$_{e}$ = 7000\,K, N$_{e}$ = 10$^{9.5}$\,cm$^{-3}$.
\\
$^{4}$Observed line ratio includes the contribution of the Ni\,{\sc ii} lines to the Si\,{\sc ii} 2335.32\,\AA\
flux. See Section 2 for details. 
\end{minipage}
\end{table*}


\begin{table*}
\centering
\begin{minipage}{140mm}
\caption{Comparison of observed and theoretical Si\,{\sc ii} line intensity ratios for $\alpha$\,Tau and the quiet Sun.} 
\begin{tabular}{lcccccc}
\hline 
Line ratio	& $\alpha$\,Tau$^1$			& Quiet Sun$^2$		& CHIANTI$^{3}$ & CLOUDY1$^{3}$ & CLOUDY2$^{3}$ 
& CLOUDY3$^{4}$ 
\\ 
\hline 
1260.42/(1264.73 + 1265.02)		& \ldots				& 0.52		& 0.51 & 0.51 & 0.50 & 0.48 
\\
1260.42/1304.37		& \ldots				& 1.90 & 1.99 & 1.83 & 1.77 & 0.22 
\\
1304.37/1309.28		& \ldots				& 0.45 & 0.58 & 0.59 & 0.55 & 0.95
\\
1526.70/1533.40		& \ldots				& 1.0 & 0.51 & 0.51 & 0.51 & 0.96 
\\
1808.01/1817.45		& 1.30		& \ldots 		& 7.90 & 7.90 & 1.93 & 0.87 
\\
1816.93/1808.01		& 2.61		& \ldots	& 1.68 & 1.68 & 2.31 & 1.03
\\
2329.23/2350.89		& 0.062 & \ldots  & 0.045 & 0.040 & 0.029 & 0.94 
\\
2335.12/2350.89		& 1.26 & \ldots & 1.20 & 1.18 & 0.75 & 0.96 
\\
2335.32/2350.89		& 3.52$^{5}$ & \ldots & 4.14 & 4.41 & 1.42 & 1.12 
\\
\hline 
\end{tabular}\label{Table:ratio2}

$^1$HST/GHRS observations of $\alpha$\,Tau from \citet{1991MNRAS.253..123J}
\\
$^{2}$HRTS observations of the quiet Sun from \citet{1992ApJ...387..726K}
\\
$^{3}$Optically thin line ratios calculated at 
T$_{e}$ = 7000\,K, N$_{e}$ = 10$^{9.5}$\,cm$^{-3}$ ($\alpha$\,Tau); T$_{e}$ = 7000\,K, N$_{e}$ = 
10$^{11}$\,cm$^{-3}$ (quiet Sun). 
\\
$^{4}$Optically thick line ratios calculated at 
T$_{e}$ = 7000\,K, N$_{e}$ = 10$^{9.5}$\,cm$^{-3}$ ($\alpha$\,Tau); T$_{e}$ = 7000\,K, N$_{e}$ = 
10$^{11}$\,cm$^{-3}$ (quiet Sun). 
\\
$^{5}$Includes contributions of Ni\,{\sc ii} lines to 2335.32\,\AA\ line flux as discussed by \citet{1991MNRAS.253..123J}.
\end{minipage}
\end{table*}


\begin{figure*}
  \centering
\hbox{
\includegraphics[width=10cm,angle=0]{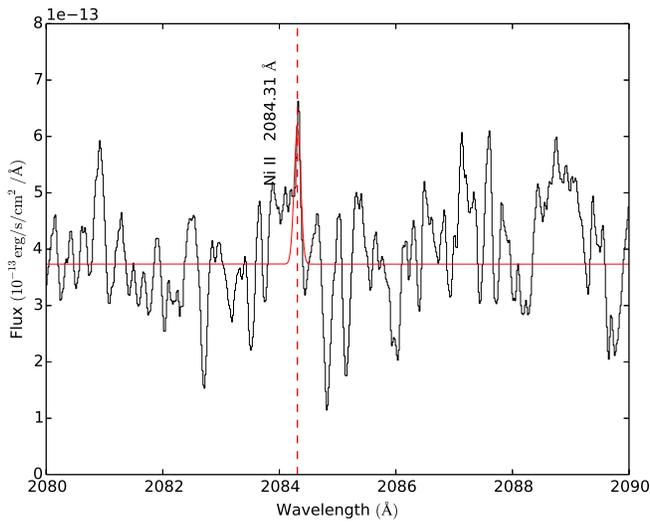} 
}
\caption{Same as Figure 1 except for the Ni\,{\sc ii} 2084.31\,\AA\ line.}
\end{figure*}


\begin{figure*}
  \centering
\hbox{
\includegraphics[width=10cm,angle=0]{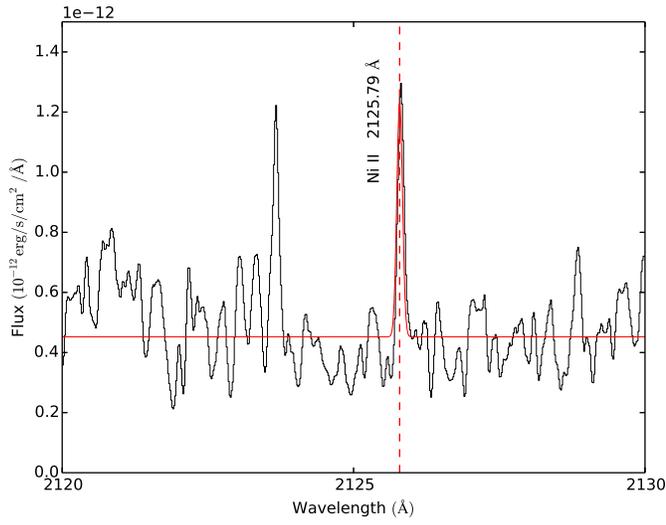} 
}
\caption{Same as Figure 1 except for the Ni\,{\sc ii} 2125.79\,\AA\ line.}
\end{figure*}


\begin{figure*}
  \centering
\hbox{
\includegraphics[width=10cm,angle=0]{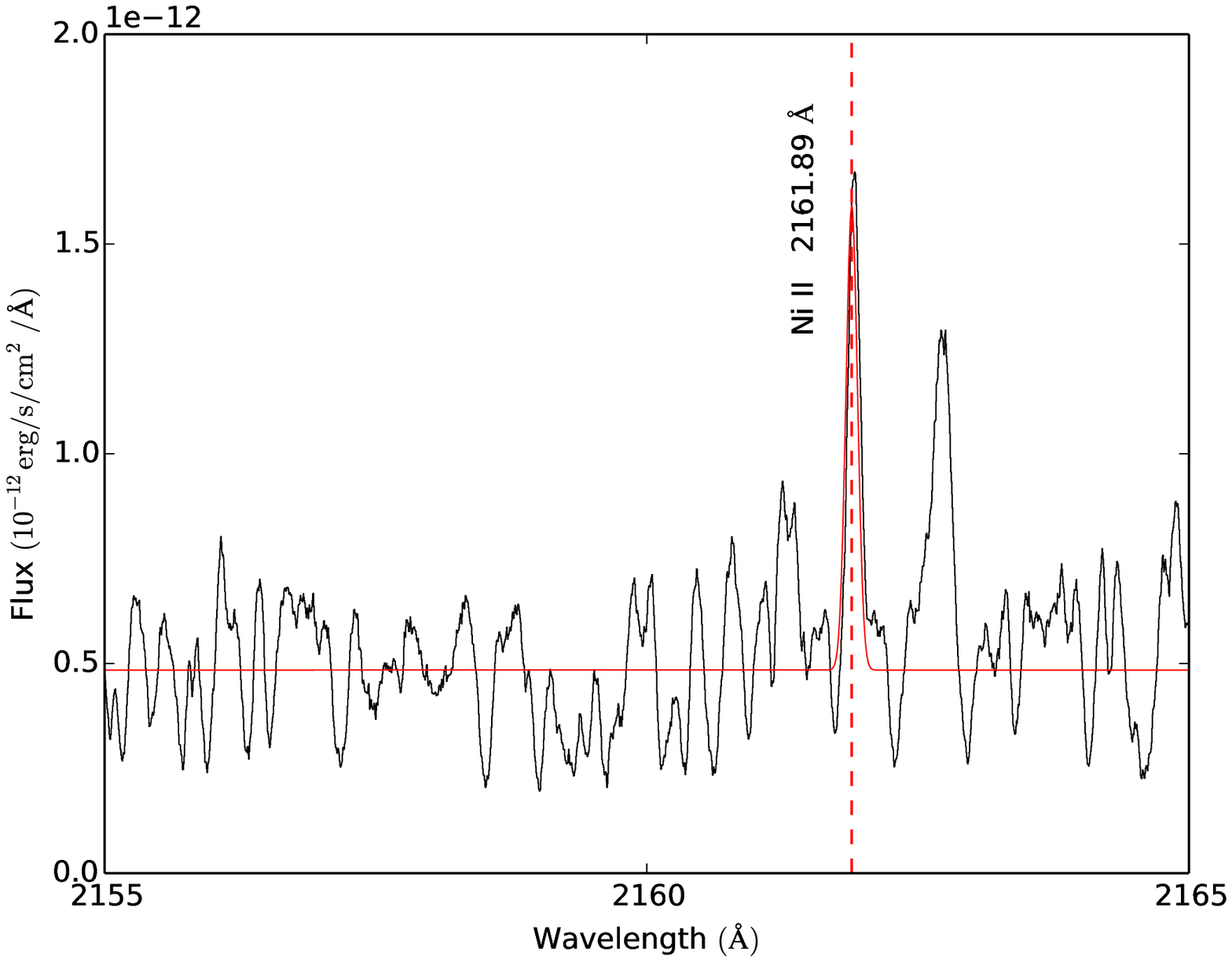} 
}
\caption{Same as Figure 1 except for the Ni\,{\sc ii} 2161.89\,\AA\ line.}
\end{figure*}


\begin{figure*}
  \centering
\hbox{
\includegraphics[width=10cm,angle=0]{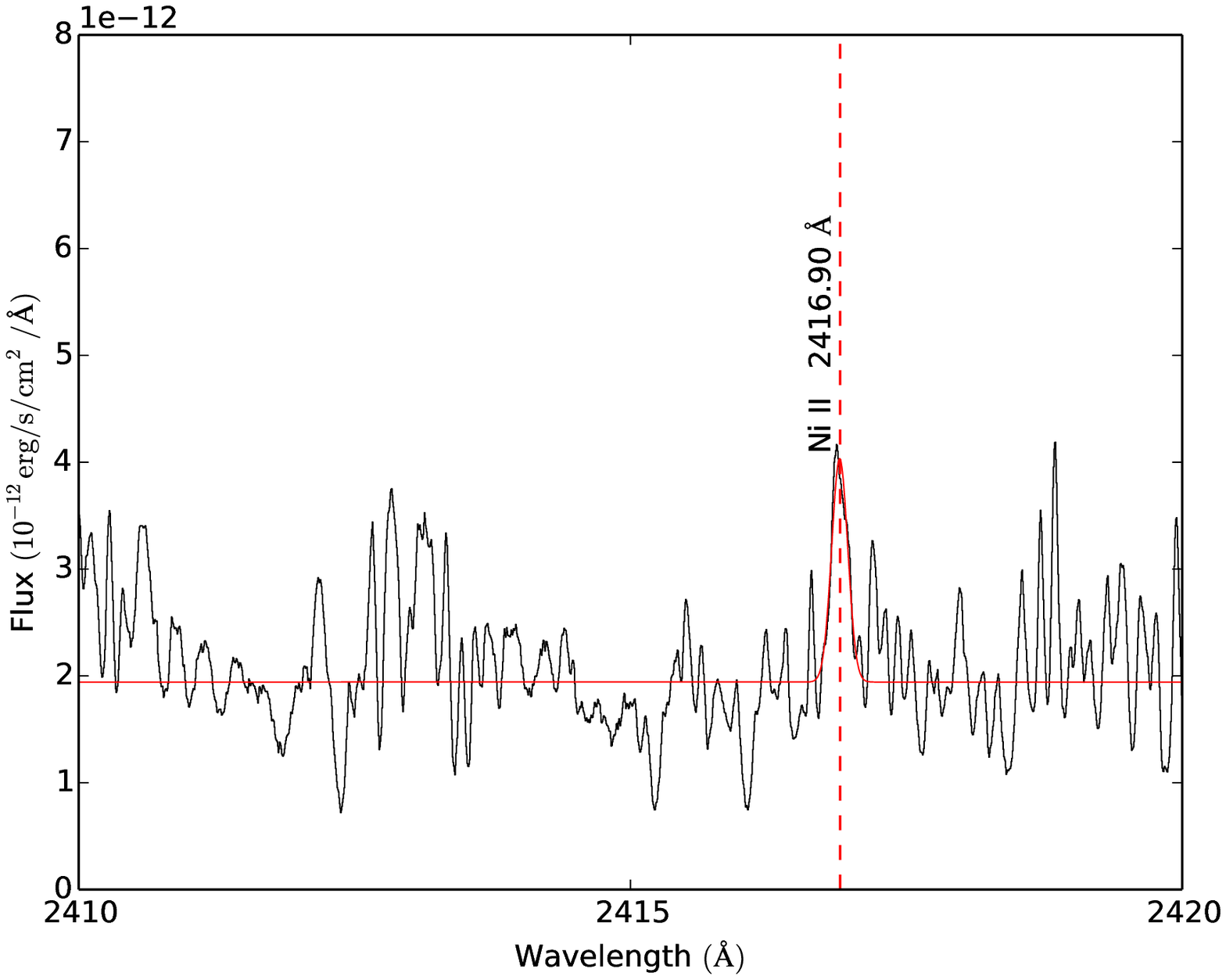} 
}
\caption{Same as Figure 1 except for the Ni\,{\sc ii} 2416.87\,\AA\ line.}
\end{figure*}


\end{document}